\documentclass[review,number,sort&compress,12pt]{elsarticle}

\usepackage{graphicx}
\usepackage{amssymb}
\usepackage{lineno}

\journal{Nuclear Instruments and Methods in Physics Research Section A}

\begin{document}
\begin{frontmatter}

\title{New air fluorescence detectors employed in the Telescope Array experiment}

\author[TITECH]{H.~Tokuno\corref{HT}}
\author[ICRR]{Y.~Tameda}
\author[ICRR]{M.~Takeda}
\author[TCU]{K.~Kadota}
\author[ICRR]{D.~Ikeda}
\author[KIN]{M.~Chikawa}
\author[OCU]{T.~Fujii}
\author[ICRR,IPMU]{M.~Fukushima}
\author[YU]{K.~Honda}
\author[SU]{N.~Inoue}
% ...

\cortext[HT]{Corresponding author. Tel.: +81-357342462, Fax: +81-357342756. 
E-mail address: htokuno@cr.phys.titech.ac.jp (H.~Tokuno).}

\address[TITECH]{Tokyo Institute of Technology, Meguro, Tokyo, Japan}
\address[ICRR]{Institute for Cosmic Ray Research, University of Tokyo, Kashiwa, Chiba, Japan}
\address[TCU]{Tokyo City University, Setagaya-ku, Tokyo, Japan}
\address[KIN]{Kinki University, Higashi Osaka, Osaka, Japan}
\address[OCU]{Osaka City University, Osaka, Osaka, Japan}
\address[IPMU]{University of Tokyo, Institute for the Physics and Mathematics of the Universe, Kashiwa, Chiba, Japan}
\address[YU]{University of Yamanashi, Interdisciplinary Graduate School of Medicine and Engineering, Kofu, Yamanashi, Japan}
\address[SU]{Saitama University, Saitama, Saitama, Japan}
% ...

\begin{abstract}
Since 2007, the Telescope Array (TA) experiment, based in Utah, USA, 
has been observing ultra high energy 
cosmic rays to understand their origins.
The experiment involves  a surface detector (SD) array and 
three fluorescence detector (FD) stations.
FD stations, installed surrounding the SD array, 
measure the air fluorescence light emitted from extensive 
air showers (EASs) for precise determination of their 
energies and species.
The detectors employed at one of the three FD stations
were relocated from the High Resolution Fly's Eye experiment.
At the other two stations, newly designed detectors were
constructed for the TA experiment.
An FD consists of a primary mirror and a camera equipped 
with photomultiplier tubes.
To obtain the EAS parameters with high accuracies, 
understanding the FD optical characteristics is important.
In this paper, we report the characteristics and installation of new 
FDs and the performances of the FD components.
The results of the monitored mirror reflectance 
during the observation time are also described in this report.
\end{abstract}


\end{frontmatter}

\linenumbers

\section{Introduction}

\begin{figure}
\begin{center}
\includegraphics[angle=0,keepaspectratio,scale=0.6]{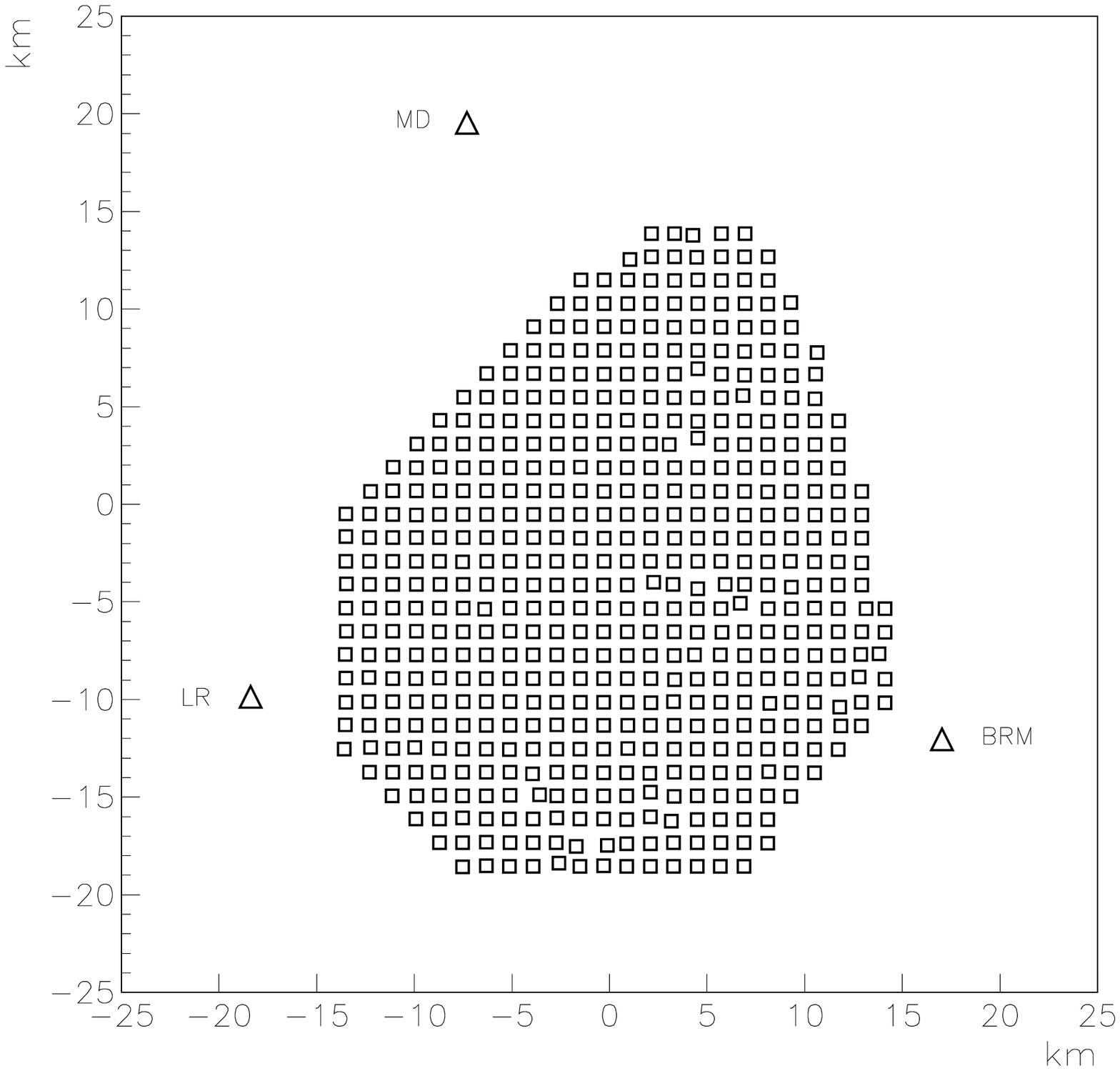}
\end{center}
\caption{Detector positions in the telescope array experiment. Squares: surface detectors (SDs), triangles: fluorescence detector (FD) telescope stations. SD array area: 700~km, SD spacing: 1.2~km, and distance between FD stations: $\sim$35~km.}\label{ta_map} 
\end{figure}

%%%%%%%%%%%%%%%%%%%%%%%%%%%%%%%%%%%%%%%%%%%%%%%%%%%%%%%%%%%%%
The Telescope Array project is a collaboration with 120 scientists 
from four nations (Japan, USA, Korea, and Russia), with the observatory 
located in Utah, USA~\cite{sogio,fukushim}.
The detectors involved in the Telescope Array experiment consist of 
surface detectors (SDs) arranged in an array and fluorescence detector (FD)
telescopes.
Fig.~\ref{ta_map} shows the detector map (squares: SD positions, 
triangles: FD station positions).
The SD array consists of 507 SDs, arranged over an area of approximately
 700~km$^{2}$ with 1.2~km spacing between the SDs~\cite{SD}.
The SDs measure the arrival timing and particle densities of extensive air 
showers (EASs) using two-layered plastic scintillators of 3~m$^2$. 
Three FD stations are placed around the SD array.
The FDs measure the air fluorescence light emitted from EASs; 
the FD data are analyzed and the longitudinal 
development of EASs is reconstructed to estimate primary energies,
arrival directions, and particle species with the aim of studying the 
nature of ultra high energy cosmic rays.
%%%%%%%%%%%%%%%%%%%%%%%%%%%%%%%%%%%%%%%%%%%%%%%%%%%%%%%%%%%%%

\begin{figure}
\begin{center}
\includegraphics[angle=0,keepaspectratio,scale=1.0]{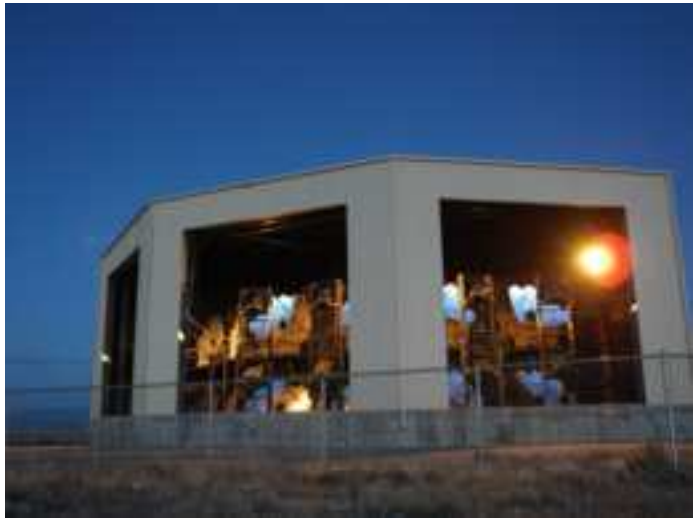}
\end{center}
\caption{Picture of the fluorescence detector station.}\label{station_image} 
\end{figure}

%%%%%%%%%%%%%%%%%%%%%%%%%%%%%%%%%%%%%%%%%%%%%%%%%%%%%%%%%%%%%
The FD station near the northwest corner of the SD array houses
fourteen FDs that were relocated from 
the High Resolution Fly's Eye (HiRes) experiment~(e.g.~\cite{HiRes}).
For these FDs and for the data obtained using these FDs,
we can use the same calibration and analysis methods 
used in the HiRes experiment~\cite{Jnm}.
The specifications, configurations, and calibrations 
of the relocated FDs can be obtained from
previous reports on HiRes (e.g.~\cite{HiRes}).
On the other hand, the southeast (BRM) and southwest 
(LR) FD sites have twelve FDs each; these FDs were newly designed 
for the TA experiment.
A picture of the FD station is shown in Fig.~\ref{station_image}.
%%%%%%%%%%%%%%%%%%%%%%%%%%%%%%%%%%%%%%%%%%%%%%%%%%%%%%%%%%%%%

%%%%%%%%%%%%%%%%%%%%%%%%%%%%%%%%%%%%%%%%%%%%%%%%%%%%%%%%%%%%%
An FD consists of a primary mirror and a 
photomultiplier tube (PMT) camera.
We employed a spherical mirror to obtain a wide field of 
view (FOV) with a sufficient focusing power. 
In addition, the size of the collecting area of the mirror was
 determined from the maximum distance for detectable EASs 
at the highest energies.
The mirror aperture and curvature radius were determined  to be
3300~mm (area of 6.8~m$^2$) and 6067~mm, respectively.
This setting realizes the detection of EASs with a sufficiently high
accuracy from a distance of 30~km and with the primary energy of 
10$^{20}$~eV.
%%%%%%%%%%%%%%%%%%%%%%%%%%%%%%%%%%%%%%%%%%%%%%%%%%%%%%%%%%%%%%

\begin{figure}
\begin{center}
\includegraphics[angle=0,keepaspectratio,scale=1.0]{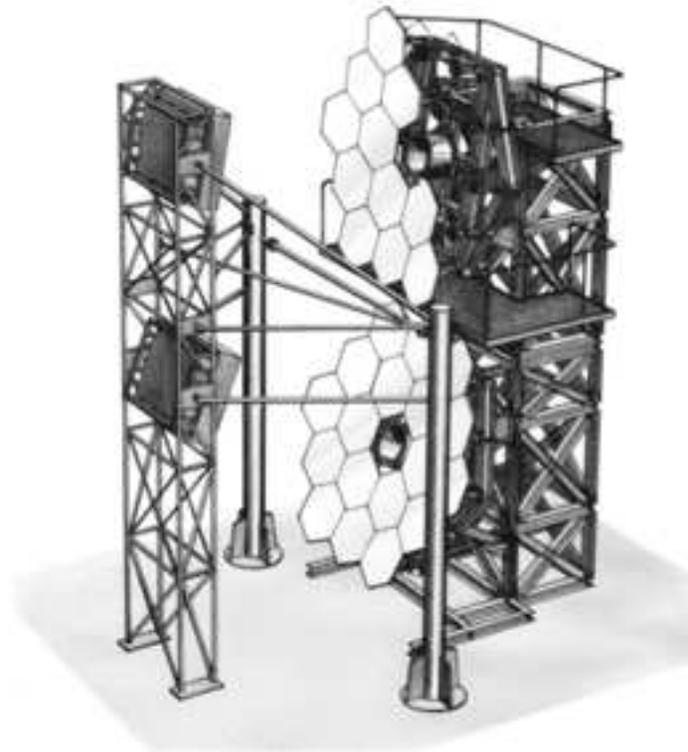}
\end{center}
\caption{Schematic view of the air fluorescence detectors. 
For visibility of mirror mounts, seven segment mirrors are 
removed from the upper telescope.}\label{telescope_image} 
\end{figure}

%%%%%%%%%%%%%%%%%%%%%%%%%%%%%%%%%%%%%%%%%%%%%%%%%%%%%%%%%%%%%%
A primary mirror is segmented into eighteen small hexagonal 
mirrors; the distance between the parallel sides of the 
hexagonal mirror was 660~mm.
A PMT camera consisting of 16$\times$16 PMTs 
and having an effective area of 860~mm $\times$ 992~mm
is set at a distance of 3000~mm from the mirror.
The FOV of each PMT is approximately 1$^{\circ}$, and 
that of the camera is 
15$^{\circ}$ in elevation and 18$^{\circ}$ in azimuth.
Fig.~\ref{telescope_image} shows a schematic view of the 
FD telescope frame equipped with a pair of telescopes,
one each in the upper and lower parts of the frame.
The heights of the upper and lower mirror centers
and the upper and lower camera centers from the floor
are 5.5, 1.5, 6.0, and 2.8~m, respectively. 
The FOV centers of the upper and lower telescopes are at elevations of
10.5$^\circ$, and 25.5$^\circ$, respectively.
As shown in Fig.~\ref{telescope_image}, no segment mirror is installed at 
the center of the primary mirror.
At this position, a jig was mounted for alignment of the optical 
system of the telescope when it was constructed.
Similarly, during normal operations, 
a UV light flasher is mounted here as a standard light 
source for the calibrations and adjustments of PMT 
gains~\cite{tokuno}.
The projected view of the FD station onto the station floor
is shown in Fig.~\ref{telescope_position}.
The station has six telescope frames;
accordingly, the FOV of a station is  from 3$^{\circ}$ to 33$^{\circ}$ in 
elevation and 108$^{\circ}$ in azimuth.
%
%%%%%%%%%%%%%%%%%%%%%%%%%%%%%%%%%%%%%%%%%%%%%%%%%%%%%%%%%%%%%

%%%%%%%%%%%%%%%%%%%%%%%%%%%%%%%%%%%%%%%%%%%%%%%%%%%%%%%%%%%%%
To evaluate the accuracies of the measured shower parameters 
such as arrival direction, primary energy, and longitudinal 
development, studies of resolutions and systematic uncertainties 
of FD telescope optics are important.
Thus, we ensured that we accurately manufactured, constructed, and installed 
the FD telescopes with utmost care and after confirmation of 
their optical characteristics.
%
%%%%%%%%%%%%%%%%%%%%%%%%%%%%%%%%%%%%%%%%%%%%%%%%%%%%%%%%%%%%%

%%%%%%%%%%%%%%%%%%%%%%%%%%%%%%%%%%%%%%%%%%%%%%%%%%%%%%%%%%%%%
%
In this paper, we describe the elements that constitute an FD 
telescope and the installation process of the telescopes.
Moreover, the reflectance of segment mirrors, which are 
continuously monitored after the installations, is also presented.
The contents of this paper are as follows.
In Sec.~\ref{camera_production} and Sec.~\ref{mirror_production},
we report the designs and productions of the PMT cameras and those of the 
segment mirrors, respectively.
Sec.~\ref{mirror_camera_installation} explains the installation of the 
cameras and mirrors.
Mirror reflectance and its variations are described
in Sec.~\ref{mirror_ref_monitor}.
%%%%%%%%%%%%%%%%%%%%%%%%%%%%%%%%%%%%%%%%%%%%%%%%%%%%%%%%%%%%%

\section{Camera production}\label{camera_production}
%%%%%%%%%%%%%%%%%%%%%%%%%%%%%%%%%%%%%%%%%%%%%%%%%%%%%%%%%%%%%
In this section, we describe the configuration of the newly 
designed PMT cameras and the results of user acceptance tests.
%%%%%%%%%%%%%%%%%%%%%%%%%%%%%%%%%%%%%%%%%%%%%%%%%%%%%%%%%%%%%

\subsection{Photomultiplier tube}

%%%%%%%%%%%%%%%%%%%%%%%%%%%%%%%%%%%%%%%%%%%%%%%%%%%%%%%%%%%%%
Each camera has 256 PMTs.
7000 PMTs including spares have been produced.
The PMT used in the TA experiment, Hamamatsu R9508 based on R6234-01, 
is specially manufactured for the experiment.
The photoelectric surface on the top of a tube is hexagonal in
shape with a distance of 60~mm between the parallel side; 
its effective area is equivalent to that of 
circle of 57~mm diameter~\cite{tokuno}.
On the bottom of the tube, a printed circuit board
of a bleeder and a preamplifier is installed.
In order to reduce the contamination of the night sky background photons 
into fluorescence signals, we use a 4~mm thick
optical filter (SCHOTT AG, BG3).
The filter is mounted on the surface of each PMT
using a self-fusing tape (Hitachi Chemical Co., Ltd., HIGHBON TAPE No.2) 
and a polyimide film tape (3M, 5434).
The transmittance of BG3 is higher than 80\% in the wavelength 
range of 305$-$395~nm, which is within the range of all the major 
air fluorescence lines~\cite{tokuno}.
%
%%%%%%%%%%%%%%%%%%%%%%%%%%%%%%%%%%%%%%%%%%%%%%%%%%%%%%%%%%%%%

%%%%%%%%%%%%%%%%%%%%%%%%%%%%%%%%%%%%%%%%%%%%%%%%%%%%%%%%%%%%%
The PMTs were inspected by the supplier before shipping and
were selected based on their cathode and anode sensitivities,
dark current, and other basic parameters.
In addition, the quantum efficiency (including collection efficiency),
and cathode uniformity were measured for sampled PMTs, once for 1000 PMTs.
We also imposed sufficiently high gains and low noise levels on PMTs 
at the experimental sites. 
The minimum required gain was 6$\times10^{4}$ with the maximum 
applied high voltage of $-$1200~V.
%%%%%%%%%%%%%%%%%%%%%%%%%%%%%%%%%%%%%%%%%%%%%%%%%%%%%%%%%%%%%

%%%%%%%%%%%%%%%%%%%%%%%%%%%%%%%%%%%%%%%%%%%%%%%%%%%%%%%%%%%%%
After installation of camera, we adjusted PMT gains to absolutely calibrated PMTs~\cite{crays} 
and measured the uniformity of the PMT response at its photoelectric surface on the sites~\cite{tokuno}.
During FD observations, each PMT gain was monitored every one hour using UV flashers.
We exchanged two of the 6144 PMTs because the gains of these two PMTs had 
decreased in the first half year. 
Since then, no PMT has been exchanged.
%due to decreasing gain in the first half year,
%and after that,  because of its failure.
%
%%%%%%%%%%%%%%%%%%%%%%%%%%%%%%%%%%%%%%%%%%%%%%%%%%%%%%%%%%%%%

\subsection{Camera}

\begin{figure}
\begin{center}
\includegraphics[angle=0,keepaspectratio,scale=0.6]{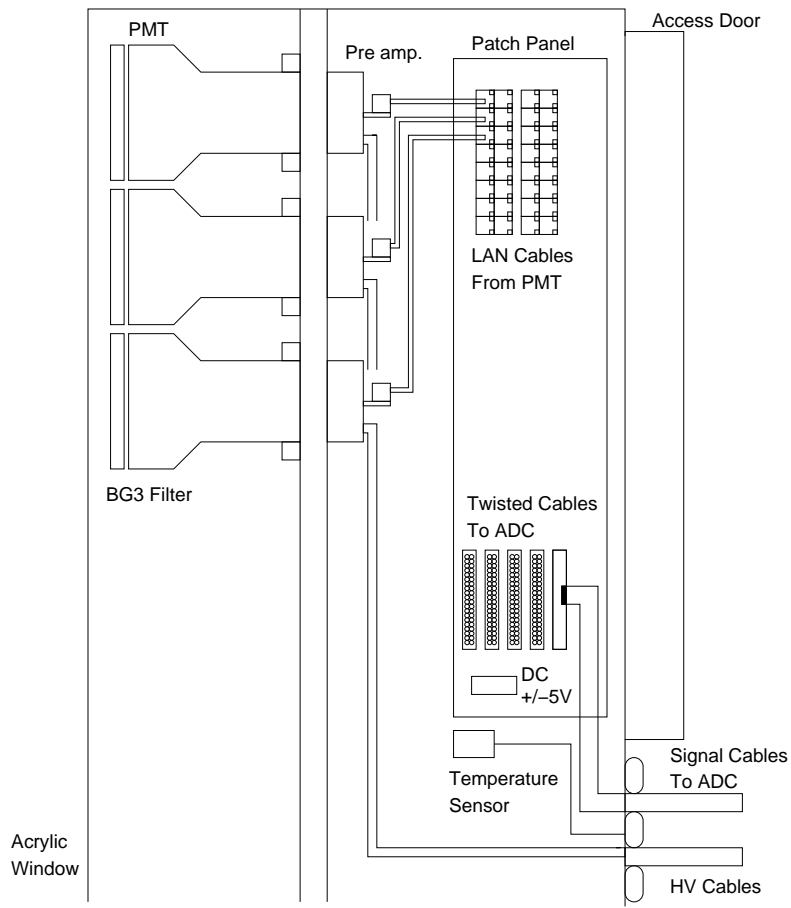}
\end{center}
\caption{Schematic side view of a PMT camera. Inside the camera, there are 256 PMTs, connected with signal and HV cables, and a temperature sensor.}\label{camera.fig} 
\end{figure}

%%%%%%%%%%%%%%%%%%%%%%%%%%%%%%%%%%%%%%%%%%%%%%%%%%%%%%%%%%%%%
A PMT camera has been placed at the primary focus of each FD telescope.
Fig.~\ref{camera.fig} shows the schematic view of the PMT camera.
The camera has a UV transparent acrylic window (KURARAY, PARAGLAS-UV00)
attached in front of PMTs for protection from dust.
The transmittance of the window is %(??kept??) 
more than 90\% in the wavelength range of 300$-$400~nm~\cite{tokuno}.
PMTs were fixed on a 16~mm thick plate of extra super duralumin
and were arranged on a triangular grid with a clearance gap of approximately
1~mm among them.
The gap geometries were measured based on the uniformity of the 
camera surface response~\cite{tokuno}.
%%%%%%%%%%%%%%%%%%%%%%%%%%%%%%%%%%%%%%%%%%%%%%%%%%%%%%%%%%%%%

\begin{figure}
\begin{center}
\includegraphics[angle=0,keepaspectratio,scale=1.0]{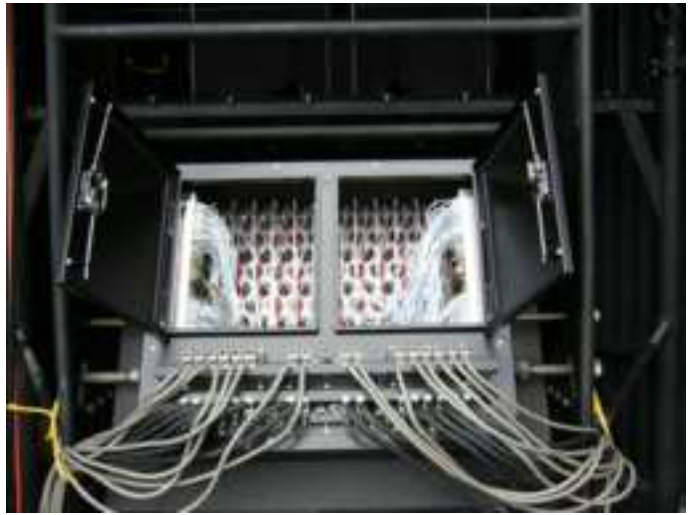}
\end{center}
\caption{Rear view of the PMT camera. }\label{camera_rear_photo.fig} 
\end{figure}

%%%%%%%%%%%%%%%%%%%%%%%%%%%%%%%%%%%%%%%%%%%%%%%%%%%%%%%%%%%%%
%
Signals from the PMTs and the DC power to the PMTs are transferred
through  shielded LAN cables of category 5 (length: 50$-$100~cm) 
between the PMTs and the patch panels inside a camera.
The patch panels connected to the LAN cables are shown 
in Fig.~\ref{camera_rear_photo.fig}.
We use a shielded 40 core twisted pair 
cable (Bando Densen Co., Ltd., BIOS-A-2820P) each for sixteen PMTs
to transfer the PMT signals from the patch panels 
in a camera to a data acquisition (DAQ) system in the electronics room 
with temperature control, situated 20~m from the camera location.
The cable length for the upper and lower cameras 
is 25.5~m and  22.3~m, respectively.
Detailed descriptions of the electronics and DAQ system are 
presented elsewhere~\cite{tameda,taketa}.
%%%%%%%%%%%%%%%%%%%%%%%%%%%%%%%%%%%%%%%%%%%%%%%%%%%%%%%%%%%%%

%%%%%%%%%%%%%%%%%%%%%%%%%%%%%%%%%%%%%%%%%%%%%%%%%%%%%%%%%%%%%
High voltages (HVs) are applied to PMTs 
via coaxial cables (Bando Densen Co., Ltd., 1.5D-2V) using 
an order-made power supply (Takasago Co., Ltd.).
The power supply has 256 outputs, and HV value
on each output can be controlled and monitored
separately through Ethernet, with the accuracy of $\pm 0.2\%$.
This uncertainty corresponds to approximately $\pm 2\%$ of the 
PMT gains under our setting.
The typical applied HV to PMTs is $-880$~V, and the rated maximum 
voltage is $-1200$~V.
As a safety measure to protect human beings, PMTs, and the power supply,
the power supply is automatically shut down when unusually high values 
are detected.
%%%%%%%%%%%%%%%%%%%%%%%%%%%%%%%%%%%%%%%%%%%%%%%%%%%%%%%%%%%%%

%%%%%%%%%%%%%%%%%%%%%%%%%%%%%%%%%%%%%%%%%%%%%%%%%%%%%%%%%%%%%
A DC power supply (KENWOOD, PW18-3AD) is connected to two patch 
panels to distribute DC voltage of $\pm 5$~V to the 256 preamplifiers, 
each connected to one PMT.
In this manner, we prepared one DC power supply for each camera.
%
%%%%%%%%%%%%%%%%%%%%%%%%%%%%%%%%%%%%%%%%%%%%%%%%%%%%%%%%%%%%%

\section{Segment mirrors and their production}\label{mirror_production}

%%%%%%%%%%%%%%%%%%%%%%%%%%%%%%%%%%%%%%%%%%%%%%%%%%%%%%%%%%%%%
In this section, we describe the specifications of the segment 
mirrors employed in the new FD telescopes of 
the TA experiment, their pre-shipment and acceptance inspections, and 
the inspection results.
%%%%%%%%%%%%%%%%%%%%%%%%%%%%%%%%%%%%%%%%%%%%%%%%%%%%%%%%%%%%%

\subsection{Segment mirrors}

%%%%%%%%%%%%%%%%%%%%%%%%%%%%%%%%%%%%%%%%%%%%%%%%%%%%%%%%%%%%%
The 3.3~m aperture spherical mirror of the FD telescope is 
segmented into eighteen small mirrors, as shown in Fig.~\ref{telescope_image}.
The mirror is made of borosilicate glass (SCHOTT AG, TEMPAX), and
each segment is hexagonal in shape with a thickness of 10.5~mm and a 
distance of 660~mm between the parallel sides.
On the surface of the mirror was deposited a 200~nm thick reflection
coating of aluminum through vacuum deposition, and on this coating
was deposited a 50~nm thick protective coating of ${\rm Al_{2}O_{3}}$ 
through  anodization process.
The wavelength that induces maximum reflectance 
depends on the thickness of the anodized coating.
This thickness was optimized to obtain the maximum 
reflectance at wavelength of 350~nm.
%
%%%%%%%%%%%%%%%%%%%%%%%%%%%%%%%%%%%%%%%%%%%%%%%%%%%%%%%%%%%%%

%%%%%%%%%%%%%%%%%%%%%%%%%%%%%%%%%%%%%%%%%%%%%%%%%%%%%%%%%%%%%
The optical parameters and fabrication accuracies of segment mirrors
depend on the  experimental requirement.
In order to optimize the optical parameters, 
we studied the reconstruction accuracies of EASs using ray-tracing, 
EAS simulation, and reconstruction programs.
From our studies, we determined the specifications of the optical parameters 
and the fabrication accuracies; the curvature
radius of the segment mirror was 6067~mm $\pm$100~mm,
the spot size at the focal point was smaller than 40~mm (it is comparable
size of the PMT surface),
and the reflectance in the wavelength range of 300~nm$-$400~nm was
 more than 85\%.
%%%%%%%%%%%%%%%%%%%%%%%%%%%%%%%%%%%%%%%%%%%%%%%%%%%%%%%%%%%%%

\subsection{Segment mirror production}

%%%%%%%%%%%%%%%%%%%%%%%%%%%%%%%%%%%%%%%%%%%%%%%%%%%%%%%%%%%%%
Segment mirrors were manufactured (Sanko Seikohjyo Co., Ltd.)
from January to November, 2004, for the BRM station 
and from March to December, 2006, for the LR station.
In all, 500 segment mirrors, including spares, were produced.
The procedure for mirror production is as follows.
First, the segment mirror was shaped by heating a planar glass on a 
ceramic model plate in a temperature-controlled electric oven.
After this spherical mirror fabrication, the mirrors were selected 
through curvature radius measurement.
Next, the mirror surface was coated with 200~nm thick aluminum 
through the vacuum deposition.
To protect the surface, 50~nm thick ${\rm Al_{2}O_{3}}$ crystal
layer was coated on it; these ${\rm Al_{2}O_{3}}$ crystals were
produced from a  solution containing 
ammonium hydroxide, tartaric acid, and ethylene glycol.
After the surface fabrication, the mirrors were selected based on
reflectance measurement.
Finally, a 150~mm diameter support disk made of the same 
material as the mirror and a stainless steel flange were bonded to 
the back of the mirror using glue (3M, Dymax 840). 
The flange was used to support the segment mirror from the telescope frame.
%
%%%%%%%%%%%%%%%%%%%%%%%%%%%%%%%%%%%%%%%%%%%%%%%%%%%%%%%%%%%%%

\subsection{Acceptance test for curvature radius and spot size}

\begin{figure}
\begin{center}
\includegraphics[angle=0,keepaspectratio,scale=0.5]{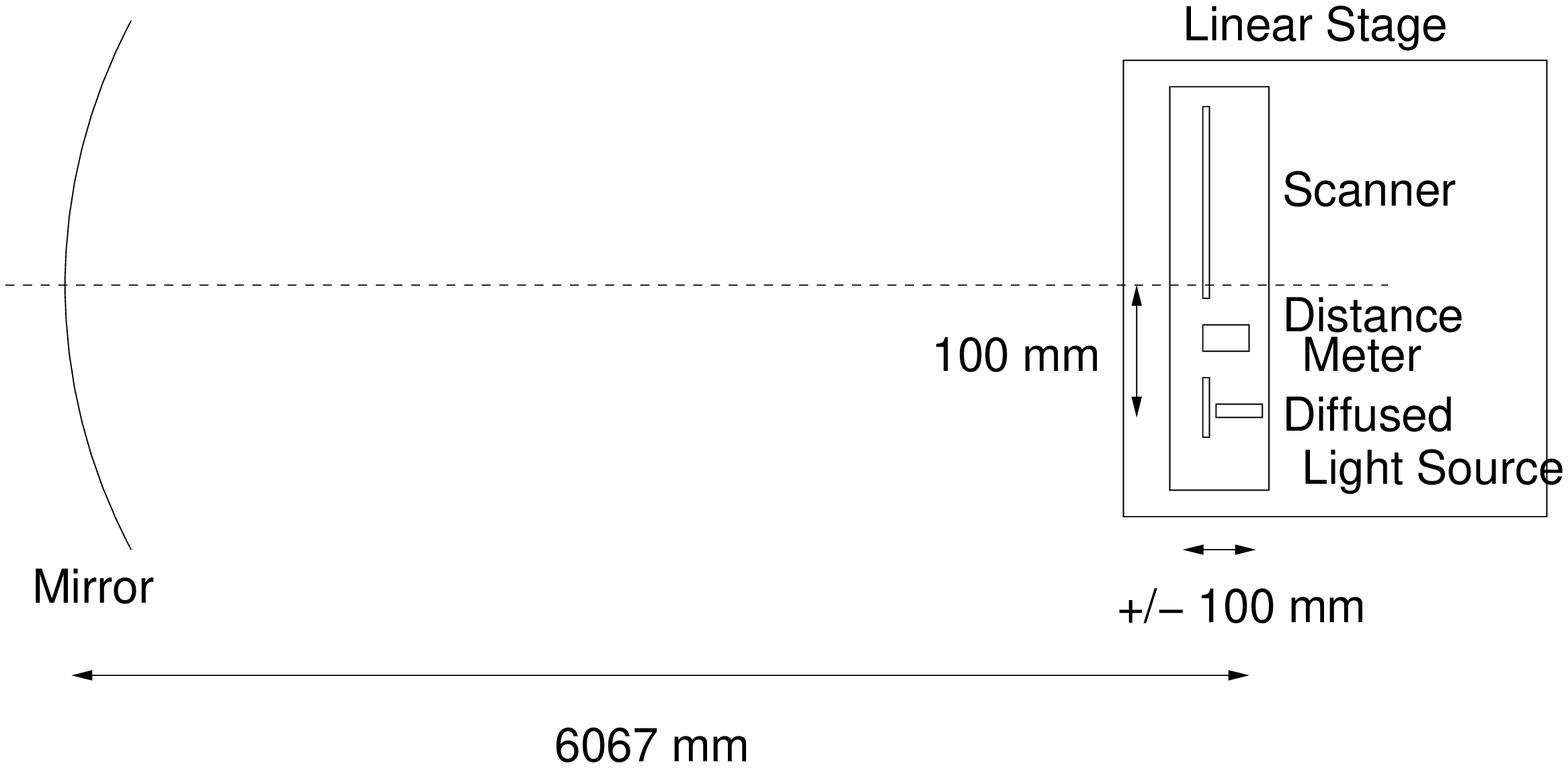}
\end{center}
\caption{Setup for the measurement of curvature radius of segment mirror.}\label{tamed_setup} 
\end{figure}

\begin{figure}
\begin{center}
\includegraphics[angle=0,keepaspectratio,scale=1.]{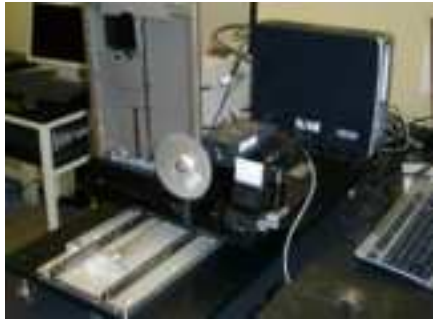}
\end{center}
\caption{Photograph of the equipment for measuring the curvature radius of segment mirror. 
This equipment includes a scanner, a distance meter, and a laser source with a diffuser plate, placed to the right on the linear stage.}\label{tamed_setup_pic} 
\end{figure}

%%%%%%%%%%%%%%%%%%%%%%%%%%%%%%%%%%%%%%%%%%%%%%%%%%%%%%%%%%%%%
For the acceptance inspections, we measured the curvature radius 
and spot size at the curvature center.
From our ray-tracing studies, we estimated an acceptable 
spot diameter of 20~mm, in which 90\% of the reflected photons fall, 
at the center of the curvature.
Because of this phenomenon, parallel incident light makes a spot
of 40~mm diameter on the camera, in which 68\% of reflected 
photons fall, in the case of a normal FD observation.
We developed a test system to measure the curvature radius of 
mirrors as the distance corresponding to the minimum spot size.
Fig.~\ref{tamed_setup} shows the schematic view of the system, and Fig.~\ref{tamed_setup_pic} shows the photograph of the same.
This system consisted of a linear precision motion stage 
(range: $\pm$250~mm, accuracy: 0.04~mm, Chuo Precision Industrial Co.,Ltd.,
MM STAGE ALS-250-C2P with controller QT-CD1); 
a diffused light source, i.e., a green laser (Kochi Toyonaka Giken Co.,Ltd., GLM-D2) 
with a diffuser plate;
an image scanner (range: 16~bit, resolution: 
2400$-$4800 dpi, Canon, CanoScan LiDE80); 
and a laser distance meter (accuracy: $\pm$1~mm, Murakami Giken, DIST pro$^4$a).
The diffused light source was set 100~mm away from the optical axis 
of the mirror, and the center of the scanner's sensitive area was fixed 
at the axisymmetrical point to the optical axis.
Both the light source and the scanner were mounted on the motion stage
and were collectively moved between 6067~$\pm$100~mm from 
the center of the mirror in 5~mm steps.
Next, we determined the curvature radius by searching for the point 
where the spot size was minimized.
From our ray-tracing calculations, we confirmed that the spot size of 
the light reflected on the scanner was minimized at the same distance 
as the curvature radius in spite of the off-axis 
alignments of the light source and the scanner.
%
%%%%%%%%%%%%%%%%%%%%%%%%%%%%%%%%%%%%%%%%%%%%%%%%%%%%%%%%%%%%%

\begin{figure}
\begin{center}
\includegraphics[angle=0,keepaspectratio,scale=1.0]{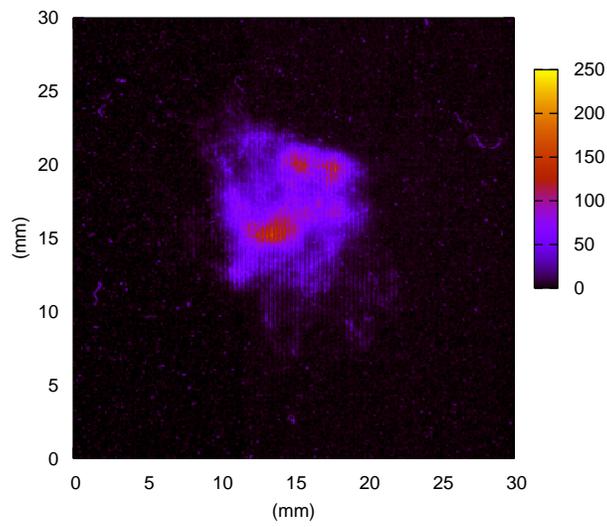}
\end{center}
\caption{Scanned image of reflected light at the distance equal to the curvature radius of the mirror. Colors show intensities of light in the electrical article.}\label{scaner_photo} 
\end{figure}

\begin{figure}
  \begin{center}
    \includegraphics[angle=0,keepaspectratio,scale=0.6]{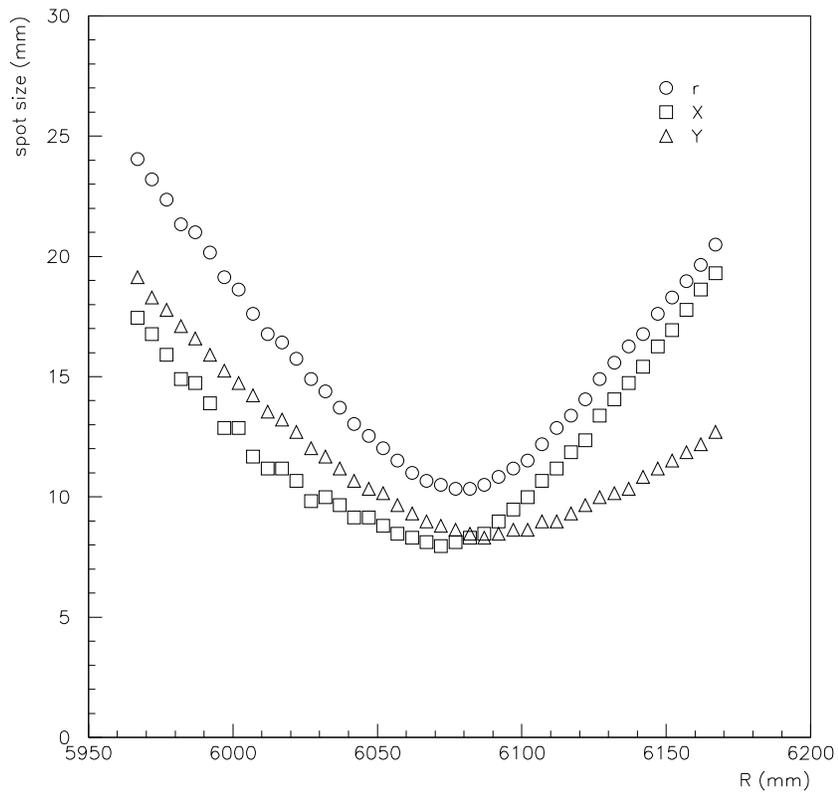}
  \end{center}
  \caption{Typical relation between spot size and distance from a segment mirror 
(circles: spot sizes; squares and triangles: projected spot sizes on X and Y axes, respectively).}\label{spotsize_distance} 
\end{figure}

\begin{figure}
  \begin{center}
    \scalebox{0.5}{\includegraphics[angle=0,keepaspectratio,scale=0.65]{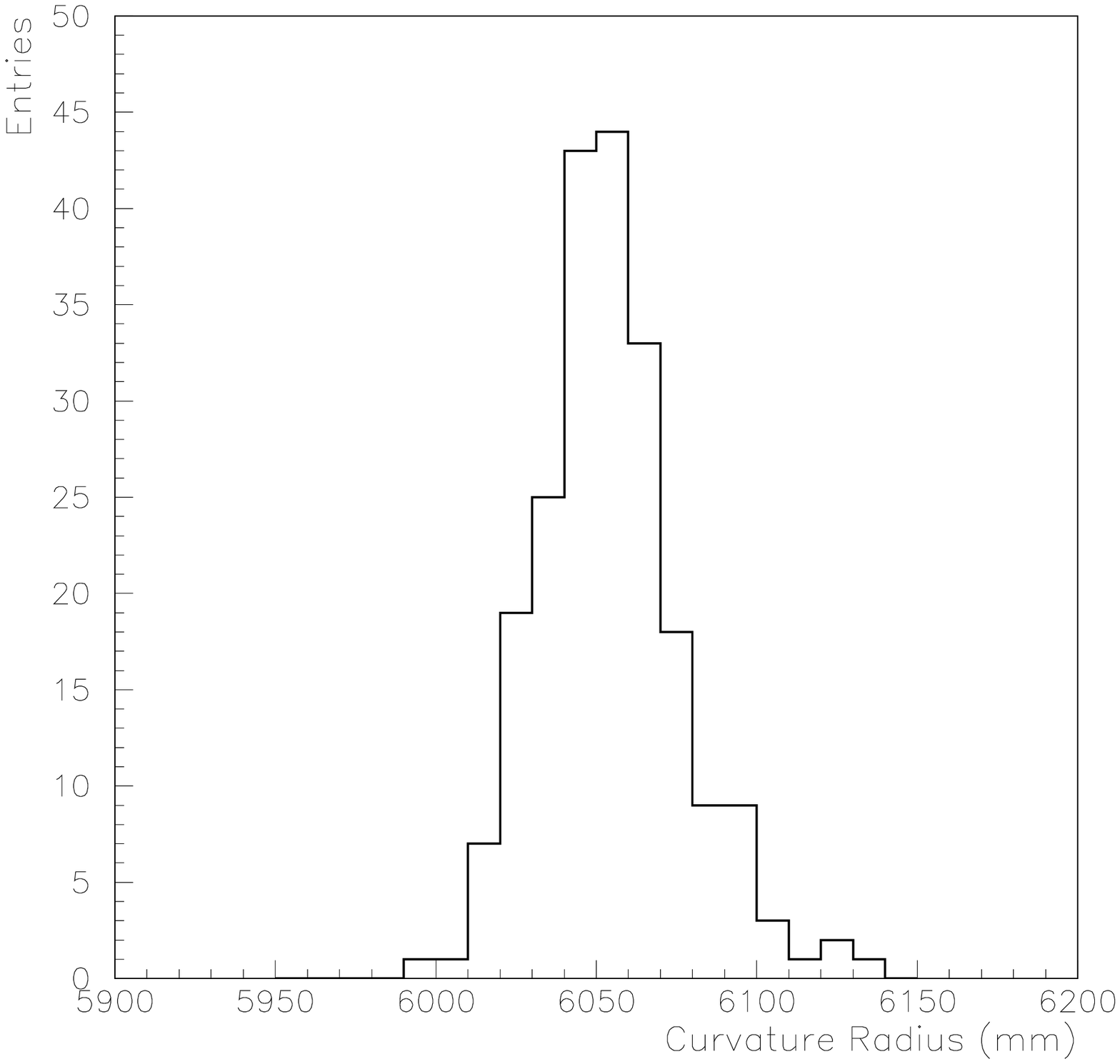}}
    \scalebox{0.5}{\includegraphics[angle=0,keepaspectratio,scale=0.65]{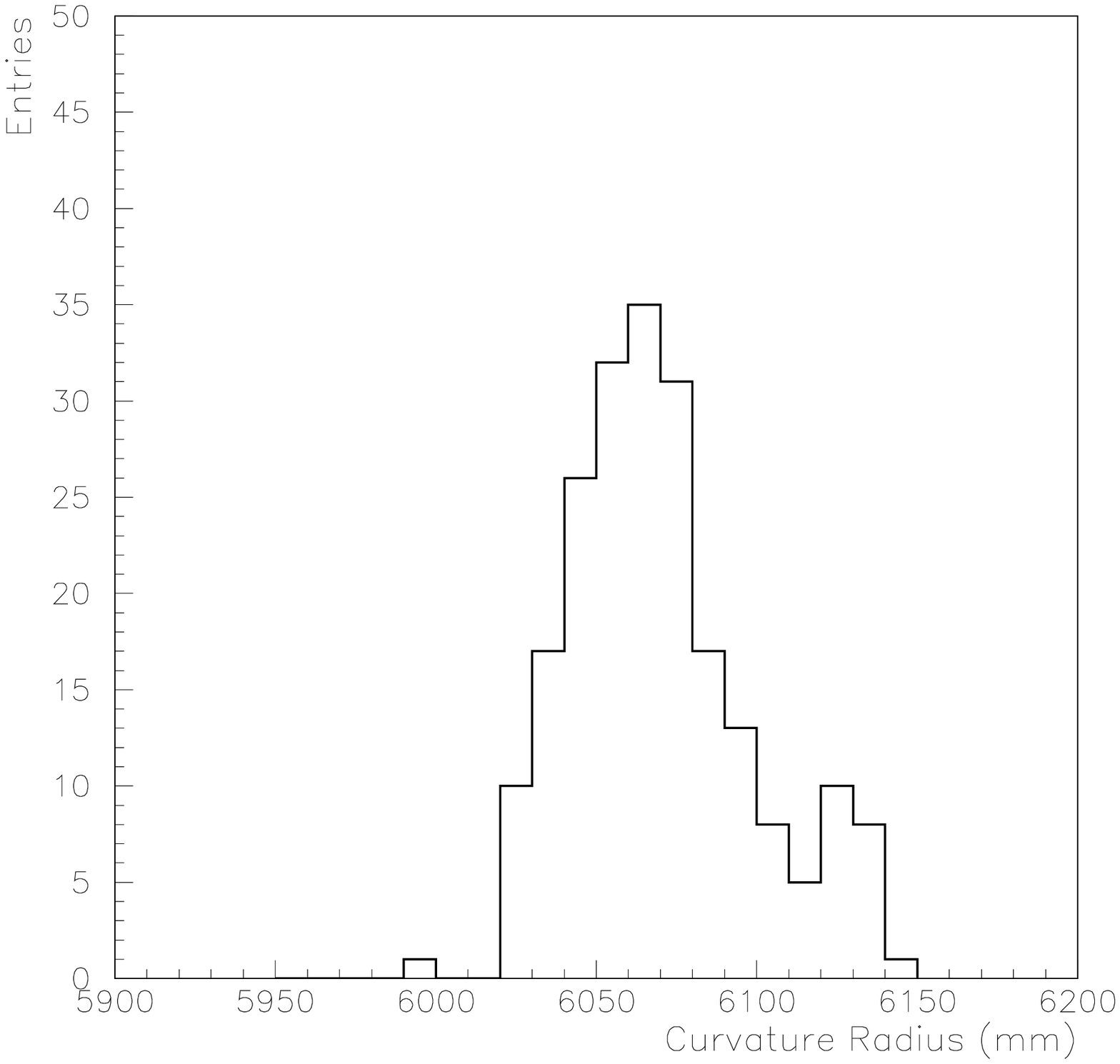}}
  \end{center}
  \caption{Distribution of the curvature radius of segment mirrors (left: BRM, right: LR).}\label{mirror_curvature_radius} 
\end{figure}

\begin{figure}
  \begin{center}
    \scalebox{0.5}{\includegraphics[angle=0,keepaspectratio,scale=0.65]{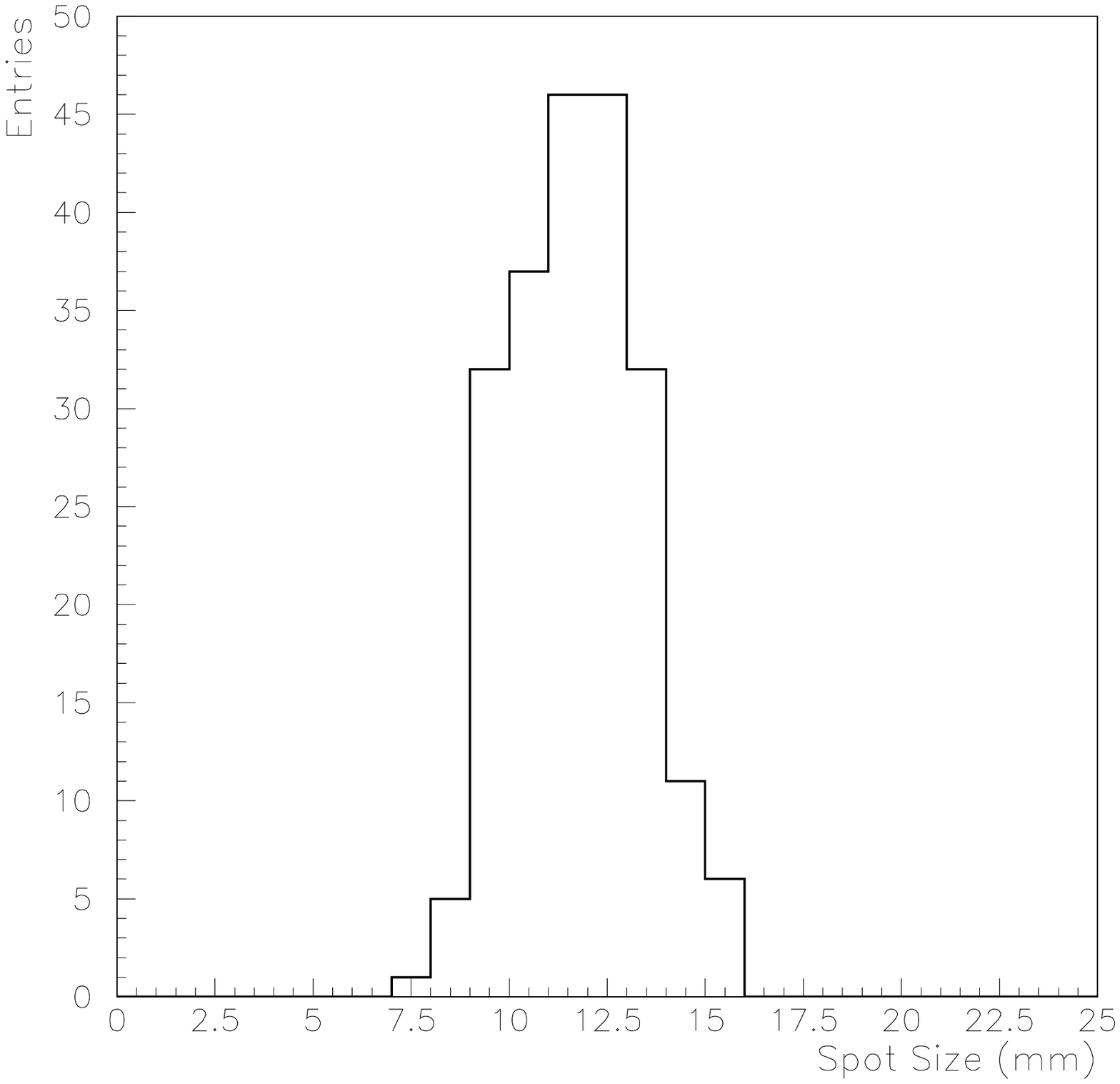}}
    \scalebox{0.5}{\includegraphics[angle=0,keepaspectratio,scale=0.65]{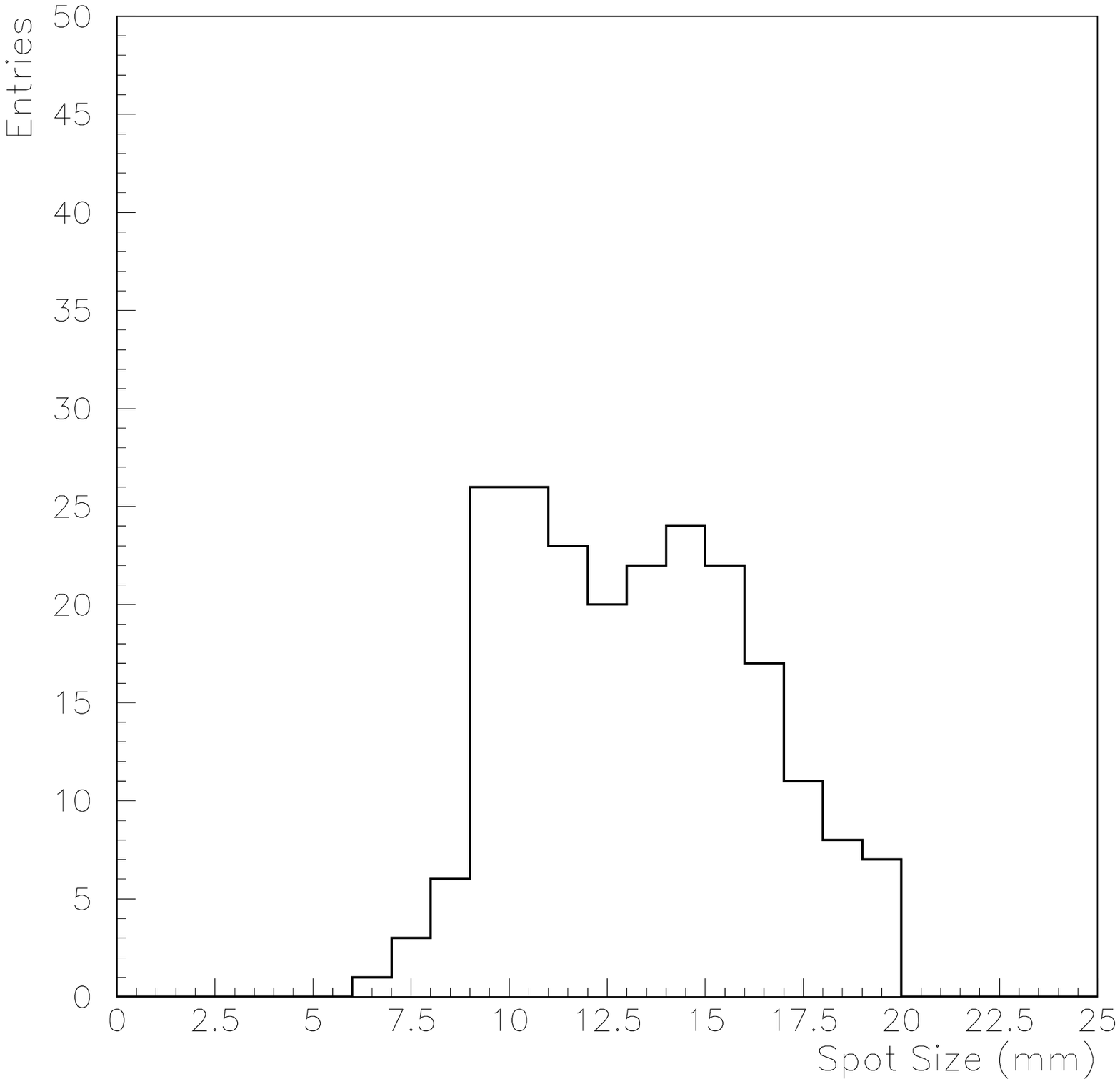}}
  \end{center}
  \caption{Distribution of the spot size of segment mirrors (left: BRM, right: LR).}\label{spotsize_at_curvature_radius} 
\end{figure}

%%%%%%%%%%%%%%%%%%%%%%%%%%%%%%%%%%%%%%%%%%%%%%%%%%%%%%%%%%%%%
A typical observed spot image is shown in Fig.~\ref{scaner_photo}.
In this analysis, the image size was defined as 
the diameter of the circle in 
which 90\% of the detected photons fall; this circle was 
centered at the weighted center of the image.
Circles in Fig.~\ref{spotsize_distance} show an example relation 
between the spot size and the distance from a mirror.
From the data points (circles) in Fig.~\ref{spotsize_distance}, 
we obtain the curvature radius to be 6082 mm.
Typically, as shown in Fig.~\ref{scaner_photo}, the image shape was 
non ideal circle, because the surface of the mirrors was 
slightly elliptical.
Accordingly, we required an additional criterion for elliptical mirrors.
We also measured spot sizes of elliptical mirrors as follows. 
First, we fitted a shot image at the beginning of a distance of 5967~mm
with a line (called X axis), and another  line (called Y axis) 
was obtained perpendicular to the X axis.
These axes were fixed during the subsequent calculations.
The shot image was projected on these axes, and the projected spot sizes
on these axes were calculated at each distance step that the 
stage away from the mirror.
Finally, we obtained the curvature radii on these axes from the 
minimum of these plots.
We also required the curvature radii on the two axes 
to be within 6067$\pm$100~mm.
The projected spot sizes on the X and Y axes for the sampled 
mirror are shown as square and triangle points, respectively, 
in Fig.~\ref{spotsize_distance}.
From this figure, the curvature radii on the X and Y axes are 
6072~mm and 6087~mm, respectively, and these radii also satisfy 
the above requirement.
Fig.~\ref{mirror_curvature_radius} shows the distribution of the 
curvature radius of the accepted mirrors, 
and Fig.~\ref{spotsize_at_curvature_radius} shows
the distribution of the normal spot size in diameter at the curvature radius.
The curvature radius of the accepted mirrors is 6057~mm (1$\sigma$: 
$-$20/$+$30~mm), and their spot size is 12~mm (1$\sigma$: $-$2/$+$3~mm)
in diameter.
%
%From now on 1~$\sigma$ means $\pm$34~\% width of distribution 
%of entries from the median value, 
%so that we have 68\% entries in the area from $-$1$\sigma$ to $+$1$\sigma$.  
Henceforth, 1~$\sigma$ indicates 68\% confidence level.
%
%%%%%%%%%%%%%%%%%%%%%%%%%%%%%%%%%%%%%%%%%%%%%%%%%%%%%%%%%%%%%

%%%%%%%%%%%%%%%%%%%%%%%%%%%%%%%%%%%%%%%%%%%%%%%%%%%%%%%%%%%%%
From the acceptance test, we found that 5\% mirrors were unacceptable.
To complement the required number of mirrors, other acceptable mirrors 
were produced additionally.
%%%%%%%%%%%%%%%%%%%%%%%%%%%%%%%%%%%%%%%%%%%%%%%%%%%%%%%%%%%%%

\subsection{Acceptance test for reflectance}
%%%%%%%%%%%%%%%%%%%%%%%%%%%%%%%%%%%%%%%%%%%%%%%%%%%%%%%%%%%%%
We employed two different acceptance tests for the segment mirror reflectance.
One is an accurate measurement in the laboratory for a wide 
wavelength range for sampled mirrors, which was performed by the manufacturer.
The other is a simple measurement with a portable spectrophotometer 
for a narrow wavelength range for all the segment mirrors.
%%%%%%%%%%%%%%%%%%%%%%%%%%%%%%%%%%%%%%%%%%%%%%%%%%%%%%%%%%%%%

\begin{figure}
  \begin{center}
    \includegraphics[angle=0,keepaspectratio,scale=0.6,trim=0 260 0 0]{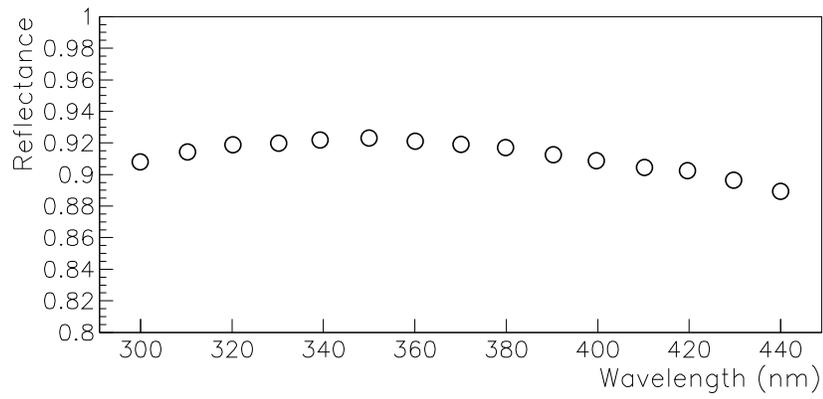}
  \end{center}
  \caption{Typical reflectance of small flat mirror, measured by the manufacturer (systematic error and measurement error are less than 1\%).}\label{mirror_ref_sanko} 
\end{figure}

%%%%%%%%%%%%%%%%%%%%%%%%%%%%%%%%%%%%%%%%%%%%%%%%%%%%%%%%%%%%%
As it is difficult to measure the reflectance of 
the curved and large segment mirrors accurately, 
the manufacturer produced small flat pieces of mirrors 
simultaneously with sampled segment mirrors and precisely 
measured the reflectance of the flat pieces as part of the
 delivery inspection process.
The reflectance of the small flat mirrors was 
measured using a spectrometer (Jasco Inc., Ubest V-550),
whose range, resolution, and accuracy are
190$-$900~nm, 2~nm, and 1\%, respectively.
Fig.~\ref{mirror_ref_sanko} shows the
typical reflectance of sampled small flat mirrors.
This reflectance is more than 90\% 
in the wavelength range of 300$-$400~nm,
containing all the major air fluorescence lines.
%
%%%%%%%%%%%%%%%%%%%%%%%%%%%%%%%%%%%%%%%%%%%%%%%%%%%%%%%%%%%%%

\begin{figure}
  \begin{center}
    \includegraphics[angle=0,keepaspectratio,scale=0.6,trim=0 260 0 0]{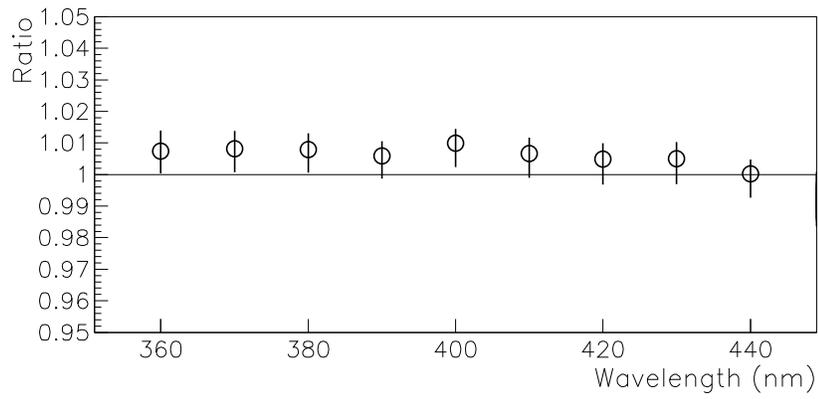}
  \end{center}
  \caption{The comparison of mirror reflectance between the NIST measurement and our measurement (plots: our measurement value divided by the NIST one, error bars: 1$\sigma$).}\label{nist_mirror_ana} 
\end{figure}

\begin{figure}
  \begin{center}
    \includegraphics[angle=0,keepaspectratio,scale=0.6,trim=0 260 0 0]{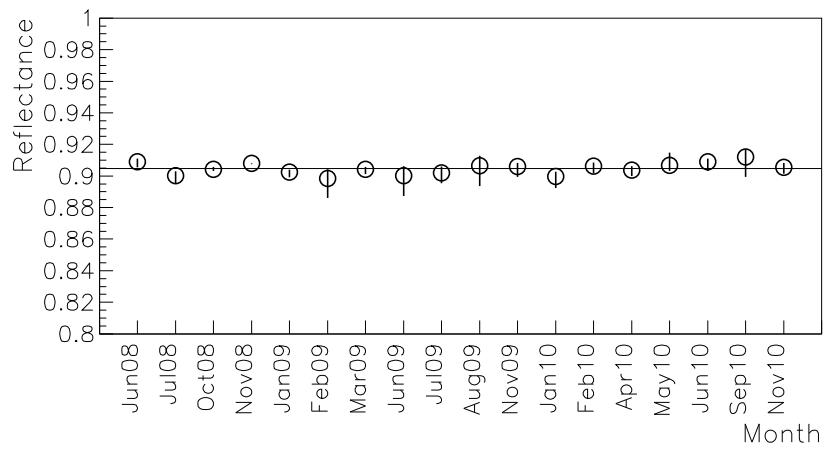}
  \end{center}
  \caption{Variation in the reflectance of a reference mirror at 360~nm from June 2008 to November 2010 (plots: median value, error bars: 1$\sigma$, horizontal line: mean value of the plots).}\label{nist_hist.fig}  
\end{figure}

%%%%%%%%%%%%%%%%%%%%%%%%%%%%%%%%%%%%%%%%%%%%%%%%%%%%%%%%%%%%%
We use a portable spectrophotometer (Konica Minolta, Inc., CM-2500d)
for acceptance inspections of all the manufactured segment mirrors 
and for routine monitoring of reflectance.
The range and its resolution are 360$-$740~nm and 10~nm, respectively.
The nominal uncertainties of measurement are less than 1\%.
To confirm the accuracy and stability of the spectrophotometer, 
we measured the reflectance of a reference mirror 
eighteen times in the period from June 2008 to   November 2010. 
The reference mirror (Ocean Optics Inc., STAN-SSH-NIST) is quartz-coated;
its reflectance was measured by the National Institute 
of Standards and Technology (NIST), USA.
Fig.~\ref{nist_mirror_ana} shows a comparison between our measurements 
and those done by NIST.
From this figure, the systematic differences are within $\pm$1\% 
and the deviations are within $\pm$1\% in the concerned wavelength range.
 Fig.~\ref{nist_hist.fig} shows the variation in the reflectance of 
the reference mirror at the wavelength of 
360~nm, measured using a  portable spectrophotometer.
The stability was 0.6\% (1$\sigma$) during the period from June 2008  
to November 2010.
%%%%%%%%%%%%%%%%%%%%%%%%%%%%%%%%%%%%%%%%%%%%%%%%%%%%%%%%%%%%%

\begin{figure}
  \begin{center}
    \includegraphics[angle=0,keepaspectratio,scale=0.6,trim=0 260 0 0]{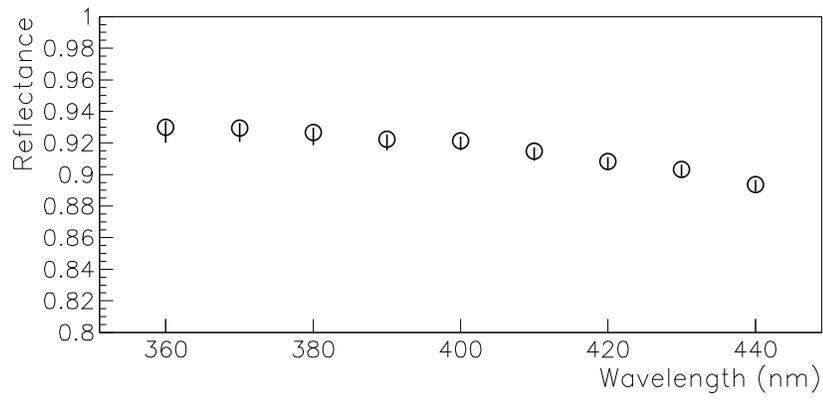}
  \end{center}
  \caption{Wavelength dependence of the reflectance of all segment mirrors in March 2007 (plots: median value of 24 primary mirrors, error bars: 1$\sigma$).}\label{mirror_ref_wavelength} 
\end{figure}

\begin{figure}
  \begin{center}
    \includegraphics[angle=0,keepaspectratio,scale=0.6]{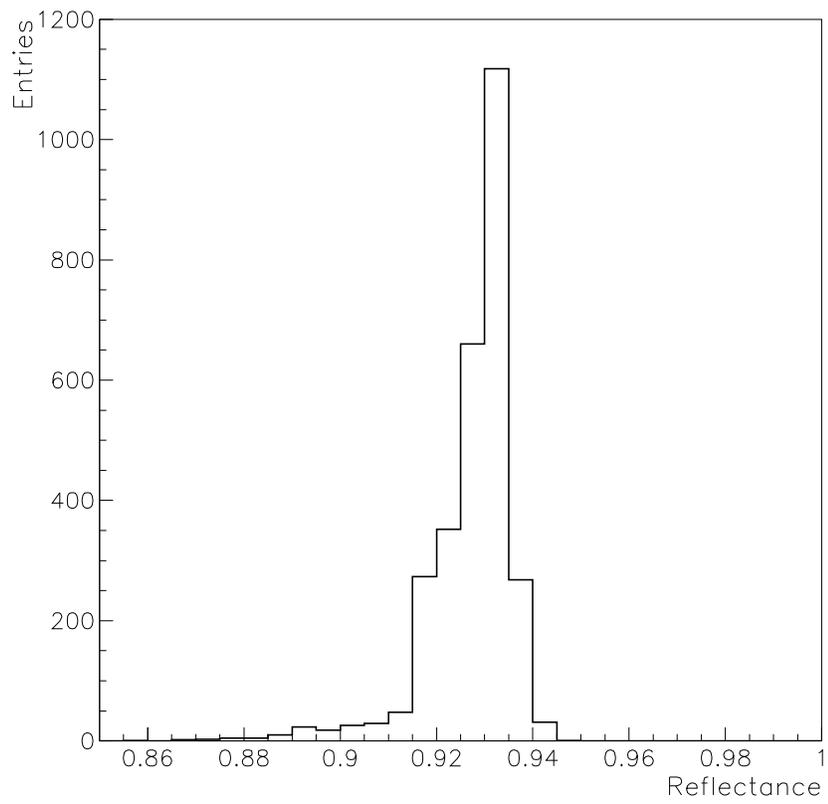}
  \end{center}
  \caption{Distribution of mirror reflectance of 24 primary mirrors at 360~nm in March 2007.}\label{mirror_ref_wavelength_360} 
\end{figure}

\begin{figure}
  \begin{center}
    \includegraphics[angle=0,keepaspectratio,scale=0.6,trim=0 260 0 0]{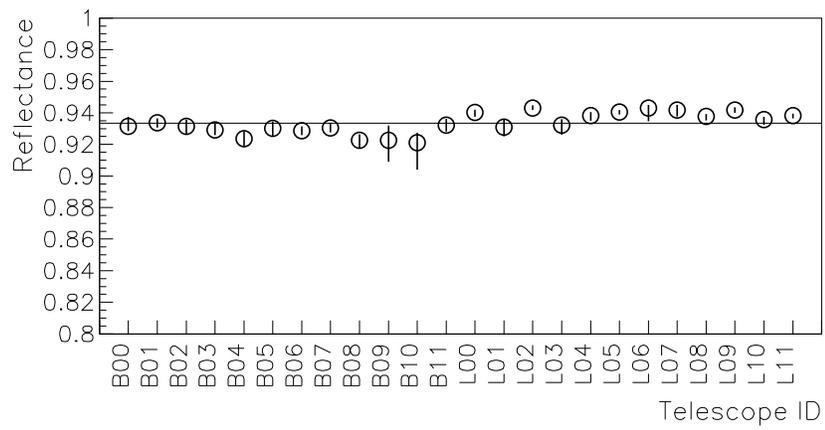}
  \end{center}
  \caption{mirror reflectance at 360~nm (horizontal axis: telescope ID, B: BRM, L: LR, plots: median value of 18 segment mirrors, error bars: 1$\sigma$, horizontal line: mean value of the plots).}\label{mirror_ref_wavelength_360_each} 
\end{figure}

%%%%%%%%%%%%%%%%%%%%%%%%%%%%%%%%%%%%%%%%%%%%%%%%%%%%%%%%%%%%%
The typical reflectance of a primary mirror composed of eighteen 
segments is shown in Fig.~\ref{mirror_ref_wavelength}.
This figure shows that the reflectance in the wavelength 
range of 360$-$400~nm is higher than 90\%.
The distribution of the reflectance at 360~nm for all the 
segment mirrors is shown in Fig.~\ref{mirror_ref_wavelength_360}.
This distribution shows that the differences in the reflectance depend 
on 108 sampling points of each primary mirror
and that the variance is less than 1\%, which is within measurement accuracies.
Fig.~\ref{mirror_ref_wavelength_360_each} 
shows the reflectance of the primary mirrors in BRM and LR stations,
which is the average of eighteen segment mirrors,
with 1$\sigma$ error bars.
All the telescopes fulfilled the requirement that the 
averaged reflectance was greater 
than 90\% in the wavelength range of 360~nm.
%%%%%%%%%%%%%%%%%%%%%%%%%%%%%%%%%%%%%%%%%%%%%%%%%%%%%%%%%%%%%

\section{Telescope installation}\label{mirror_camera_installation}

%%%%%%%%%%%%%%%%%%%%%%%%%%%%%%%%%%%%%%%%%%%%%%%%%%%%%%%%%%%%%
Mirrors and cameras were shipped from Japan to our observatory in Utah, USA, and these items were installed at BRM 
from February 2005 to July 2006 and at LR in March 2007.
In this section, 
we describe the installations of the FD telescopes and 
their alignment.
%%%%%%%%%%%%%%%%%%%%%%%%%%%%%%%%%%%%%%%%%%%%%%%%%%%%%%%%%%%%%

\subsection{Installation of telescope frame}

\begin{figure}
  \begin{center}
\includegraphics[angle=0,keepaspectratio,scale=0.4]{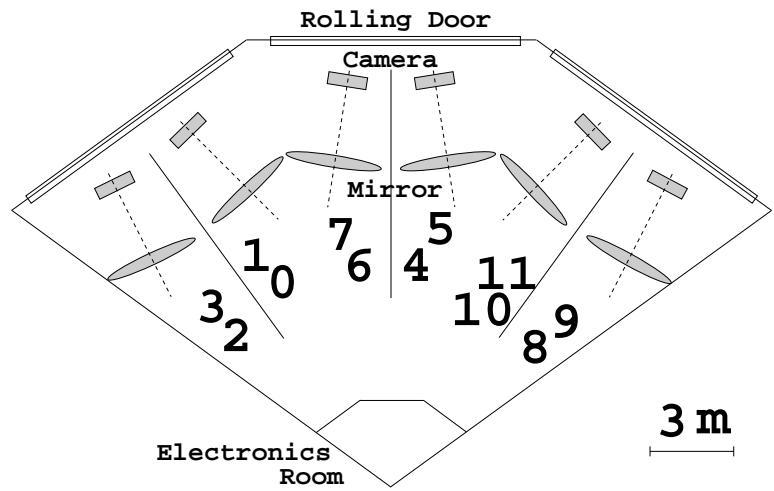}
  \end{center}
  \caption{Schematic projected view of the FD station onto the floor (solid lines: datum lines, 
    dashed lines: azimuthal lines of each telescope frame). 
    %The FOV direction of the neighboring telescope is alternated. 
    The telescopes are numbered according 
    to their FOVs, as shown in this figure; the lower (upper) telescopes have even (odd) numbers.
  }\label{telescope_position} 
\end{figure}

%%%%%%%%%%%%%%%%%%%%%%%%%%%%%%%%%%%%%%%%%%%%%%%%%%%%%%%%%%%%%
As shown in Fig.~\ref{telescope_position},
six telescope frames were installed in each station (BRM and LR)
with each frame having two telescopes.
Hence, in all, twelve telescopes were installed in each station.
%
%%%%%%%%%%%%%%%%%%%%%%%%%%%%%%%%%%%%%%%%%%%%%%%%%%%%%%%%%%

%%%%%%%%%%%%%%%%%%%%%%%%%%%%%%%%%%%%%%%%%%%%%%%%%%%%%%%%%%
A station building has three rolling doors
to protect the telescopes from sunlight, 
rain, wind, and other natural conditions.
The telescopes are directly exposed to air during FD observation,
because no other protection windows were installed.
%%%%%%%%%%%%%%%%%%%%%%%%%%%%%%%%%%%%%%%%%%%%%%%%%%%%%%%%%%

%%%%%%%%%%%%%%%%%%%%%%%%%%%%%%%%%%%%%%%%%%%%%%%%%%%%%%%%%%
When the wind speed was higher than 15~m/s, doors were closed
to prevent themselves from being stuck.
We required less than 0.1$^{\circ}$ FOV deflections of 
the telescope with wind speeds lower than 15~m/s.
Wind-induced deflections of the telescopes were studied 
by the manufacturer (Mitsui Engineering \& Shipbuilding Co., Ltd.)
using simulation programs and by applying wind-tunnel tests on a 
structure model of a station including telescopes. 
From their studies, winds from the side of the telescope 
building induce the maximum deflections in the telescope frame. 
Because winds pass through the building from one door to another,
their effective speed is not reduced by the building.
%Because the winds pass through from a door to other doors 
%without effective speed reducing by the building.
%
However, the current design of the frame fulfills our requirement 
with regard to frame deflections, even in the worst condition.
Based on the studies of the manufactures, the design of telescope frame structure was fixed.
%
%
%%%%%%%%%%%%%%%%%%%%%%%%%%%%%%%%%%%%%%%%%%%%%%%%%%%%%%%%%%%%%

%%%%%%%%%%%%%%%%%%%%%%%%%%%%%%%%%%%%%%%%%%%%%%%%%%%%%%%%%%%%%
To install the telescope frames precisely, we conducted a survey and drew 
datum lines and points for telescope geometry. 
Using a gyro-compass, the absolute azimuthal direction 
was measured with reference to a mountain's peak.
Uncertainties of the gyro-compass measurement was 1$^{\prime}$, which was 
confirmed from the star observation using the gyro-compass.
%from
Three datum lines, shown as solid lines in Fig.~\ref{telescope_position},
were drawn on the floor with relative accuracies of 1$^{\prime}$
between these lines.
Azimuthal lines of each telescope frame, shown as dashed 
lines in Fig.~\ref{telescope_position}, were drawn from 
the datum lines with accuracies of 0.1$^\circ$ using a 
measuring tape.
From the azimuthal lines, reference points for the telescope 
frames were marked with accuracies of 1~mm.
%
%%%%%%%%%%%%%%%%%%%%%%%%%%%%%%%%%%%%%%%%%%%%%%%%%%%%%%%%%%%%%

%%%%%%%%%%%%%%%%%%%%%%%%%%%%%%%%%%%%%%%%%%%%%%%%%%%%%%%%%%%%%
We installed the telescope frames, which can be separated into
frames for cameras and mirrors, as shown 
in Fig.~\ref{telescope_image}, in accordance with the following procedures.
The telescope frames were placed and fixed on the reference points,
and before installing real cameras, a prototype standard camera
was attached to the camera frames.
This prototype standard camera had the same weight and size of  
a real camera and was equipped with reference lines on the surface for 
adjusting positions.
A steel cylinder with a diameter of 500~mm is mounted at 
the center of each primary mirror, as shown in 
Fig.~\ref{telescope_image}.
The central cylinder was used to adjust the alignment of segment 
mirrors and the camera of the telescope.
%%%%%%%%%%%%%%%%%%%%%%%%%%%%%%%%%%%%%%%%%%%%%%%%%%%%%%%%%%%%%

%%%%%%%%%%%%%%%%%%%%%%%%%%%%%%%%%%%%%%%%%%%%%%%%%%%%%%%%%%%%%
%
Telescope directions were adjusted by the following procedures.
First, we installed a laser range meter on the central cylinder.
Next, we adjusted the elevation angle to the specified value,
monitored using a digital tilt meter, (Digital Protractor, Pro 360),
with an accuracy of $\pm$0.2$^{\circ}$.
We adjusted the azimuthal angle of the cylinder for the central axis
laser to point the string of a plumb 3000~mm away from the cylinder 
set exactly above the reference line on the floor.
Then, the position of the camera frame was precisely adjusted with 
reference to the central axis laser.
The upper limit for the installed error on the distance between
the camera and the cylinder is $\pm$30~mm.
However, this misalignment was resolved when we installed mounts 
of segment mirrors whose 
heights were adjusted to remove the distance error.
On the other hand, to reduce the construction cost, 
we required relatively low 
accuracies of the direction of the camera, which are within 1$^{\circ}$.
This affects the negligible uncertainties of the effective area of the 
camera, the order of which is $\cos(1^{\circ})$.
After the adjustment of the camera position, the prototype standard 
camera was unmounted from the camera frame.
%%%%%%%%%%%%%%%%%%%%%%%%%%%%%%%%%%%%%%%%%%%%%%%%%%%%%%%%%%%%%%%%

\begin{figure}
  \begin{center}
    \includegraphics[angle=0,keepaspectratio,scale=1.5]{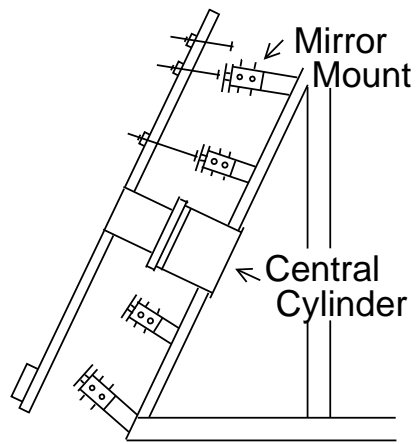}
  \end{center}
\caption{Schematic view of the equipment used for adjusting segment mirror mount. The equipment was mounted on the central cylinder of the mirror.}\label{mirror_setup} 
\end{figure}

%%%%%%%%%%%%%%%%%%%%%%%%%%%%%%%%%%%%%%%%%%%%%%%%%%%%%%%%%%%%%%%%
Finally, the positions of the segment mirror mount were adjusted.
Fig.~\ref{mirror_setup} shows the schematic view of a jig called 
``BANANA1''
for adjusting the positions of the segment mirror mount.
BANANA1 was mounted on the central cylinder of mirrors 
and can be rotated
around the central axis of the cylinder.
The equipment had two arms, one of which had three protruding legs
of different lengths; 
the other arm had a counter weight.
By rotating BANANA1, we adjusted the length of every segment mirror 
mount for one of the three legs on BANANA1
to be aligned with the reference point on the mount surface.
At the same time, the direction of each segment mirror was
 roughly adjusted using BANANA1. 
%
%%%%%%%%%%%%%%%%%%%%%%%%%%%%%%%%%%%%%%%%%%%%%%%%%%%%%%%%%%%%%

\subsection{Installation of cameras and mirrors}

\begin{figure}
\begin{center}
\includegraphics[angle=0,keepaspectratio,scale=0.6]{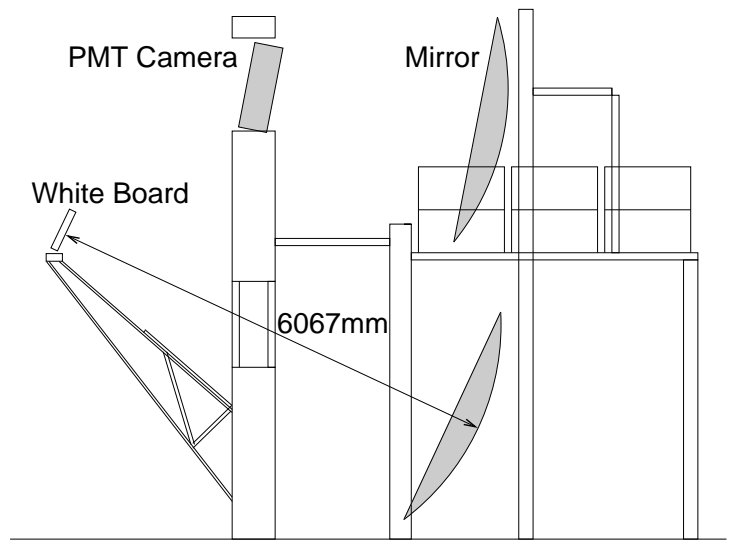}
\end{center}
\caption{Schematic view of the equipment for adjusting 
segment mirror direction. 
The equipment is mounted on a camera support frame. A white board with LEDs is placed at a distance same as the mirror curvature radius of 6067~mm using this equipment.}\label{mirror_setup2} 
\end{figure}

\begin{figure}
  \begin{center}
    \includegraphics[angle=0,keepaspectratio,scale=0.5]{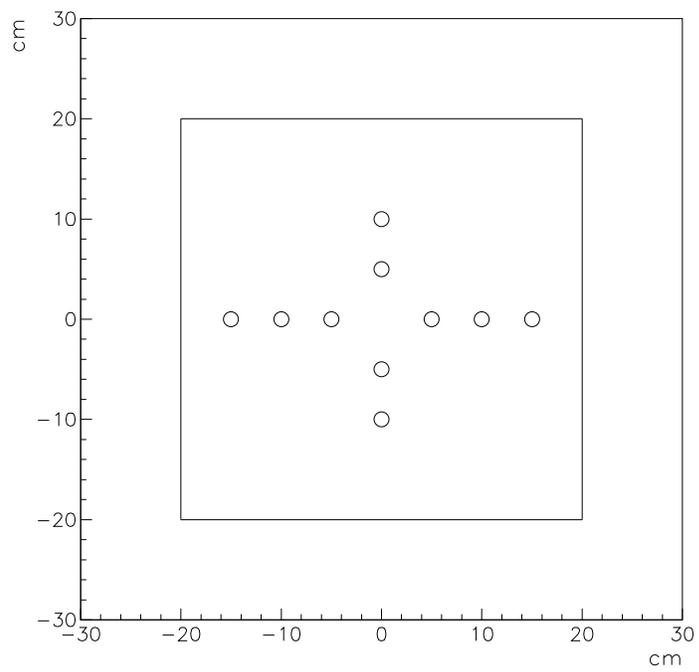}
  \end{center}
\caption{Schematic view of LED positions on the white board of BANANA3.}\label{banana3_face} 
\end{figure}

%%%%%%%%%%%%%%%%%%%%%%%%%%%%%%%%%%%%%%%%%%%%%%%%%%%%%%%%%%%%%
The cameras and mirrors were set on the telescope frame 
by the following procedures.
We set the segment mirrors on their mounts and precisely aligned each
direction by screwing in and out two adjusting bolts on a mount.
To adjust the direction of segment mirrors, we use an equipment 
called ``BANANA3'', shown in Fig.~\ref{mirror_setup2}.
The equipment was mounted on the camera frame, and a white board was 
perpendicularly set on the optical axis of the primary mirror.
The distance between segment mirrors and the board 
was 6067~mm, which is the same distance as the curvature radius of the mirrors.
Green LEDs were connected to the white board, 5, 10, and 15~cm away 
from the center of curvature, as shown in Fig.~\ref{banana3_face}.
When the direction of the segment mirror was adjusted,
the LED lights were reflected to the symmetrical point 
about the center of curvature. 
The direction was adjusted with an accuracy of less than 
$~0.1^{\circ}$,
which was estimated from the accuracy of center position 
determination of reflected light of 5~mm (1/4 diameter
of the reflected lights)
and from the distance between mirror and the white board, 6067~mm.
For parallel lights, this setting accuracy provided 
us the spots of 40~mm diameter, in which 68\% of reflected 
photons fall, at the camera surface.
After these adjustments, the segment mirror mounts were locked up.
%
%%%%%%%%%%%%%%%%%%%%%%%%%%%%%%%%%%%%%%%%%%%%%%%%%%%%%%%%%%%%%

%%%%%%%%%%%%%%%%%%%%%%%%%%%%%%%%%%%%%%%%%%%%%%%%%%%%%%%%%%%%%
PMT cameras were mounted on the camera frames with 
an accuracy of 10~mm on the camera plane.
This accuracy corresponds to $\sim$0.2$^{\circ}$ uncertainty 
of the telescope's pointing direction.
PMT cameras were connected by signal, HV, and DC cables.
Finally, the  wiring was confirmed through test operations.
%%%%%%%%%%%%%%%%%%%%%%%%%%%%%%%%%%%%%%%%%%%%%%%%%%%%%%%%%%%%%

%%%%%%%%%%%%%%%%%%%%%%%%%%%%%%%%%%%%%%%%%%%%%%%%%%%%%%%%%%%%%
In all, the directions of the telescope are adjusted with a 
few 0.1$^\circ$ accuracies.
These directions were confirmed with a star light analysis, which 
uses a similar method as in reference~\cite{hires_star}.
Our FDs record air fluorescence signals with background photons (BGPs),
as we employ a DC coupling for the PMT signal readout.
Thus, we obtain the variations in BGP with a normal trigger rate 
of 2$-$3~Hz.
When a star traverses the FOV of a PMT, the number of BGPs increases 
and decreases in a few minutes.
The time variations of BGPs are simulated from 
star positions and telescope directions.
Using these estimated time variations of BGPs, we studied
the telescope directions and spot sizes of the primary mirror.
The correction values of the FOV direction of FD obtained from this study 
are used in our simulation and the EAS reconstruction programs.
Our results of the star light analysis will be reported 
in a future paper.
%%%%%%%%%%%%%%%%%%%%%%%%%%%%%%%%%%%%%%%%%%%%%%%%%%%%%%%%%%%%%

\section{Mirror reflectance monitoring}\label{mirror_ref_monitor}

\begin{figure}
\begin{center}
\includegraphics[angle=0,keepaspectratio,scale=0.6,trim=0 260 0 0]{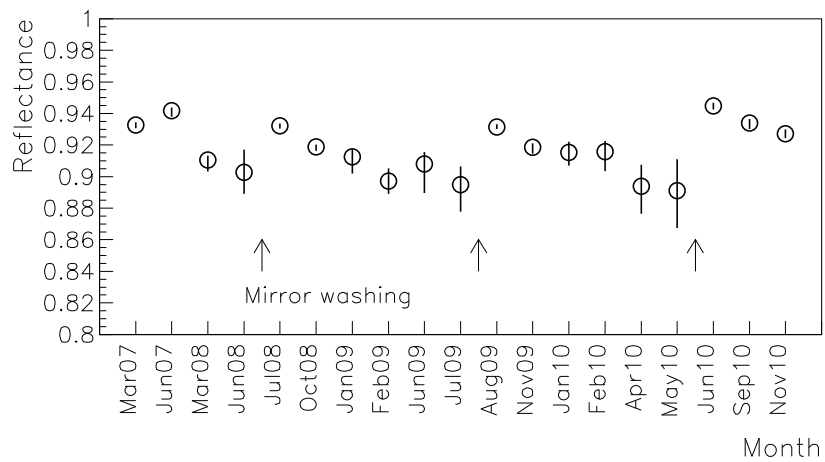}
\end{center}
\caption{Variation in the mirror reflectance at 360~nm of a typical lower telescope LR04 (plots: median value, error bars: 1$\sigma$). The mirror was washed after these measurements in  July 2008, August 2009, and May 2010.}\label{mirror_ref_time_dep} 
\end{figure}

%%%%%%%%%%%%%%%%%%%%%%%%%%%%%%%%%%%%%%%%%%%%%%%%%%%%%%%%%%%%%
The reflectance of the segment mirrors is monitored every months.
Mirror reflectance decreases with time, 
because our FDs are exposed to the air directly and because
dust is  attached to the mirrors during observations.
In addition, the reflectance of lower mirrors decreases more 
rapidly than that of upper mirrors.
Fig.~\ref{mirror_ref_time_dep} shows the time variations of 
the averaged mirror reflectance at 360~nm of a typical lower 
telescope LR04.
%
%It is found that the decreasing in winter is larger than that 
%in summer from this figure.
%
The largest decrement of 4~\% occurred between June 2007 and March 2008.
When we analyze the FD data from that period, we take this 4~\% 
as the systematic uncertainty of the mirror reflectance.
%
%%%%%%%%%%%%%%%%%%%%%%%%%%%%%%%%%%%%%%%%%%%%%%%%%%%%%%%%%%%%%

\begin{figure}
\begin{center}
\includegraphics[angle=0,keepaspectratio,scale=0.6,trim=0 260 0 0]{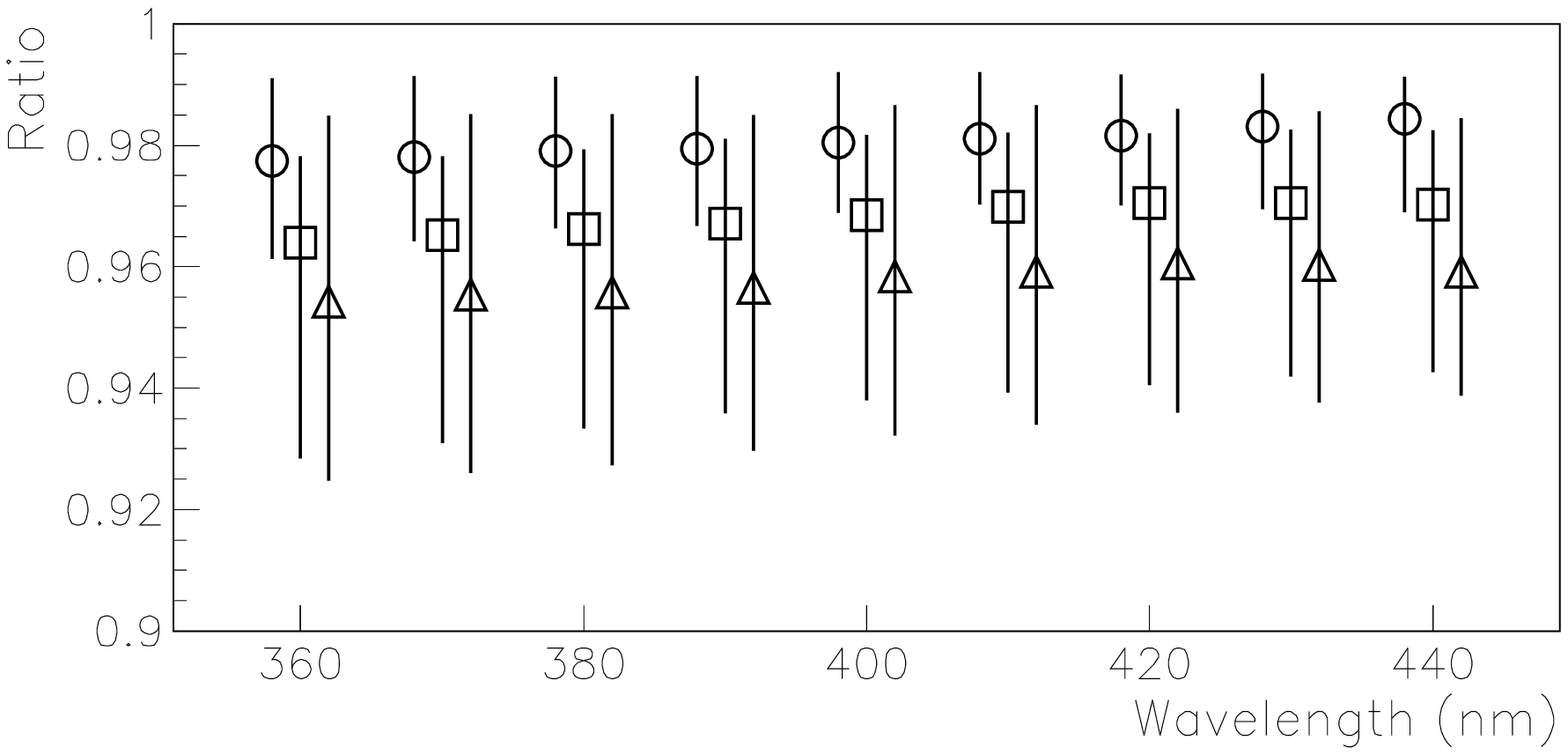}
\end{center}
\caption{Differences in mirror reflectance before and after mirror washing at each wavelength (ratio: before/after, plots: median value of 24 telescopes, error bars: 1$\sigma$, circles: 2008, squares: 2009, triangles: 2010). 
These error bars show the non-uniformity of reflectance just before mirror washing.
For visibility, the circles and triangles are slightly moved on the horizontal axis. }\label{ref_diff} 
\end{figure}

%%%%%%%%%%%%%%%%%%%%%%%%%%%%%%%%%%%%%%%%%%%%%%%%%%%%%%%%%%%%%
Every summer, the mirrors are washed using purified water and 
detergent for sensitive equipments (Alconox, Inc., LIQUI-NOX).
Purified water is obtained from tap water filtered through 
charcoal and deionization filters.
After washing, the mirrors are rinsed using purified water 
and dried naturally.
Fig.~\ref{mirror_ref_time_dep} shows that the reflectance 
of the mirror after washing is recovered to the same level as in
 the first installation.
Fig.~\ref{ref_diff} shows the differences in mirror reflectance 
before and after washing at each wavelength; these differences 
have no clear wavelength dependencies.
Based on this result, we mainly monitor the reflectance at 360~nm.
%
%%%%%%%%%%%%%%%%%%%%%%%%%%%%%%%%%%%%%%%%%%%%%%%%%%%%%%%%%%%%%

\section{Summary}

%%%%%%%%%%%%%%%%%%%%%%%%%%%%%%%%%%%%%%%%%%%%%%%%%%%%%%%%%%%%%
In Japan, we have produced PMT cameras and mirrors for use in the 
air fluorescence 
detector of the Telescope Array experiment.
The cameras and mirrors that passed our acceptance inspections were 
shipped to our experimental site in Utah, USA.
The telescope frames, mirrors, and cameras were installed 
at the FD stations.
The directions of the telescope are adjusted with an accuracy of 
a few 0.1$^\circ$, and the directions are confirmed with a star light 
analysis.
These parameters are used in our detector simulation and the 
EAS reconstruction programs.
Monitoring of the performances of the cameras and mirrors begin according 
to the following schedules: the gains of PMT camera, hour by hour during 
FD operation; mirror reflectance, every few months.
Our mirrors are washed every summer, and their reflectance after washing 
is recovered to the same level as in the first installation.
%%%%%%%%%%%%%%%%%%%%%%%%%%%%%%%%%%%%%%%%%%%%%%%%%%%%%%%%%%%%%

\section*{Acknowledgments}
The Telescope Array experiment is supported 
by the Japan Society for the Promotion of Science through
Grants-in-Aid for Scientific Research on Specially Promoted Research (21000002) 
``Extreme Phenomena in the Universe Explored by Highest Energy Cosmic Rays'', 
and the Inter-University Research Program of the Institute for Cosmic Ray 
Research;
by the U.S. National Science Foundation awards PHY-0307098, 
PHY-0601915, PHY-0703893, PHY-0758342, and PHY-0848320 (Utah) and 
PHY-0649681 (Rutgers); 
by the National Research Foundation of Korea 
(2006-0050031, 2007-0056005, 2007-0093860, 2010-0011378, 2010-0028071, R32-10130);
by the Russian Academy of Sciences, RFBR
grants 10-02-01406a and 11-02-01528a (INR),
IISN project No. 4.4509.10 and 
Belgian Science Policy under IUAP VI/11 (ULB).
The foundations of Dr. Ezekiel R. and Edna Wattis Dumke,
Willard L. Eccles and the George S. and Dolores Dore Eccles
all helped with generous donations. 
The State of Utah supported the project through its Economic Development
Board, and the University of Utah through the 
Office of the Vice President for Research. 
The experimental site became available through the cooperation of the 
Utah School and Institutional Trust Lands Administration (SITLA), 
U.S.~Bureau of Land Management and the U.S.~Air Force. 
We also wish to thank the people and the officials of Millard County,
Utah, for their steadfast and warm support. 
We gratefully acknowledge the contributions from the technical staffs of our
home institutions and the University of Utah Center for High Performance Computing
(CHPC). 
We thank HAMAMATSU Photonics K.K., Sanko Seikohjyo Co., Ltd.,
Quality Steel Fabricating And Welding, Inc, T\&D Maintenance, Nitto Kogyo Corporation, and  Mitsui Engineering \& Shipbuilding Co., Ltd. for their kind support.

\clearpage
\newpage

\clearpage
\newpage

\clearpage
\newpage

\clearpage
\newpage

\clearpage
\newpage

\clearpage
\newpage

\clearpage
\newpage

\clearpage
\newpage

\clearpage
\newpage

\clearpage
\newpage

\clearpage
\newpage
%/ta/home/htokuno/ta_work_user_htokuno/work_htokuno/mirror_nim/tamed/hist2.kumac

\clearpage
\newpage
%/ta/home/htokuno/ta_work_user_htokuno/work_htokuno/mirror_nim/tamed/hist2.kumac

\clearpage
\newpage
%/ta/home/htokuno/ta_work_user_htokuno/work_htokuno/mirror_ref/for_nim/hist10.kumac

\clearpage
\newpage
%/ta/home/htokuno/ta_work_user_htokuno/work_htokuno/mirror_nim/data/nist_mirror_ref.kumac

\clearpage
\newpage
%/ta/home/htokuno/ta_work_user_htokuno/work_htokuno/mirror_ref/for_nim/hist12.kumac

\clearpage
\newpage
%/ta/home/htokuno/ta_work_user_htokuno/work_htokuno/mirror_ref/for_nim/hist9.kumac

\clearpage
\newpage
%/ta/home/htokuno/ta_work_user_htokuno/work_htokuno/mirror_ref/for_nim/hist.kumac

\clearpage
\newpage
%/ta/home/htokuno/ta_work_user_htokuno/work_htokuno/mirror_ref/for_nim/hist2.kumac

\clearpage
\newpage

\clearpage
\newpage

\clearpage
\newpage
%/home/htokuno/work/tex/ta_mirror_nim_rev2/banana3

\clearpage
\newpage
%/ta/home/htokuno/ta_work_user_htokuno/work_htokuno/mirror_ref/for_nim/hist14.kumac

\clearpage
\newpage
%/ta/home/htokuno/ta_work_user_htokuno/work_htokuno/mirror_ref/for_nim/figure/hist11.kumac


\begin{thebibliography}{100}
\bibitem{sogio} H. Kawai  et al., Nucl. Phys. B Proc. Suppl, 175-176, (2008) 221
\bibitem{fukushim} H. Kawai  et al.,  J. Phys. Soc. Jpn. Suppl. A, 78 (2009) 108
\bibitem{HiRes} T. Abu-Zayyad et al., NIM A, 450 (2000) 253
\bibitem{SD} T. Nonaka et al., Proc. of 30th ICRC, 5 (2007) 1005, and now preparing a detailed paper 

\bibitem{Jnm} J. N. Matthews et al., Proc. of 30th ICRC, 5 (2007)  1157
\bibitem{tokuno} H. Tokuno et al., NIM A, 601 (2009) 364
\bibitem{crays} H. Tokuno et al., Proc. of 30th ICRC, 5 (2007) 1013, and now preparing a detailed paper 
\bibitem{tameda} Y. Tameda et al., NIM A, 609 (2009)  227
\bibitem{taketa} A. Taketa et al., Proc. of 29th ICRC, 8 (2005) 209, and now preparing a detailed paper
\bibitem{hires_star} P. A. Sadowski et al., Astropart. Phys, 18 (2002) 237
%\bibitem{Auger} J. Abraham et al., 2010, NIM A, v620, 227 ?? 
%\bibitem{star_catalog} G.I. Thompson et al., Catalog of Stellar Ultraviolet Fluxes, The Science Research Council (1978).
%\bibitem{star_catalog2} http://heasarc.gsfc.nasa.gov/W3Browse/td1/td1.html
\end{thebibliography}
\end{document}